\documentclass[]{jd04}    
\usepackage[]{graphicx}

\Title[Evolution of G--M Main Sequence Stars]{Evolution over Time of
Magentic Dynamo Driven UV \& X-ray Emissions of dG-M Stars
and Effects on Hosted Planets}
\Author{Guinan$^1$, Edward F.}
\Author{Engle$^1$, Scott G.}
\Address{$^1$Department of Astronomy \& Astrophysics\\
Villanova University, Villanova, PA 19085, USA\\}{}

\ShortAuthors{Guinan, E.F.}


\begin{document}
\maketitle

\begin{abstract}
The evolution over time of the magnetic activity and the resulting X-ray
and UV coronal and chromospheric emissions of main-sequence dG, dK, and
dM stars with widely different ages are discussed. Young cool stars
spin rapidly and have correspondingly very robust magnetic dynamos and
strong coronal and chromospheric X-ray -- UV (XUV) emissions. However,
these stars spin-down with time as they lose angular momentum via magnetized
winds and their magnetic generated activity and emissions significantly
decrease. Studies of dK--dM stars over a wide range of ages and rotations
show similar (but not identical) behavior. Particular emphasis is
given to discussing the effects that XUV emissions have on the atmospheres
and evolution of solar system planets as well as the increasing number of
extrasolar planets found hosted by dG--dM stars. The results from modeling
the early atmospheres of Venus, Earth and Mars using recently determined
XUV irradiances and inferred winds of the young Sun are also briefly
discussed. For example, the loss of water from juvenile Venus and Mars
can be explained by action of the strong XUV emissions and robust winds
of the young Sun. We also examine the effects of strong X-ray and UV
coronal and chromospheric emissions (and frequent flares) that dM stars
have on possible planets orbiting within their shrunken habitable zones
(HZs) -- located close to the low luminosity host stars (HZ $\leq$ 0.4 AU).
Dwarf M stars make interesting targets for further study because of
their deep outer convective zones (CZs), efficient dynamos, frequent flares
and strong XUV emissions. Furthermore, a large fraction of dM stars
are very old ($>$5 Gyr), which present intriguing possibilities for
the development of highly advanced modes of intelligent life on planets
that may orbit them.
\end{abstract}

\vspace{-2mm}
\section{Introduction}
\vspace{-2mm}
  
Over the last 15 years, as part of the ``Sun in Time'' program, we have
been carrying out an in-depth study of solar-type stars with different
ages, rotation rates and correspondingly vastly different levels of magnetic
activity. The main goals of this program are to investigate the solar
dynamo, and to determine the dependence of coronal X-ray/EUV emissions
and chromospheric/transition region (TR) FUV/UV emissions on stellar age
and rotation. The latter goal was expanded to include the contruction
of XUV spectral irradiance tables for the Sun over its main sequence
lifetime (e.g. Guinan et al. 2003; Ribas et al. 2005). These spectral
XUV irradiances are very important
in studies of the effects of the young Sun's increased radiation on
paleoplanetary atmospheres and environments. We use a homogeneous sample
of $\sim$15 single, nearby dG0--5
stars as proxies for the Sun (and solar-type stars), covering ages from
$\sim$100 Myr to 8.5
Gyr. The ages of the younger program stars are inferred from memberships
in various clusters / moving groups, while the ages for the older stars
are secured from isochronal fits. The rotation periods are obtained chiefly
from ground-based photometry (star spot modulation) or from chromospheric
Ca {\sc ii} rotationally induced variability. Excellent relationships
are found among rotation, age and activity (see Guinan et al. 2003). The
program stars have
been observed from X-ray--UV using {\em ROSAT}, {\em ASCA}, {\em Chandra},
{\em XMM}, {\em EUVE}, {\em FUSE}, {\em IUE} and {\em HST}. 

The strength of our sample is its homogeneity. In
essence, we use a $\sim$1 M$_\odot$ star as a laboratory to study the effects
of the stellar dynamo (with similar convective zone (CZ) depths) by
varying the only free parameter, P$_{\rm rot}$.
Our studies have shown that the Sun's magnetic field has steadily
declined as its rotation slowed due to magnetic breaking. Also, the
study of the young Sun's (as well as other dG stars) XUV fluxes
using solar proxies reveals unprecedented diagnostics for the state of
the younger solar system and the physics of the much more active early
Sun (Ayres et al. 1996; Guinan et al. 2003; Ribas et al. 2005). Fig.
1 shows a plot of rotation period vs.
stellar age for most of our ``Sun in Time'' sample, with a power law fit,
showing the spin-down of the Sun and other solar-type stars
with age. Fig. 2 shows the related decrease in X-ray luminosity (L$_{\rm x}$)
with
age for the ``Sun in Time'' sample. As shown in these figures, the young
Sun was rotating over 10x faster than the present Sun and had
correspondingly high (5--1000x) levels of XUV coronal and chromospheric
emissions.
Most of the observed ranges in L$_{\rm x}$ for the older, less active
stars shown in Fig. 2 arise from solar-like activity cycles.
Some results of the ``Sun in Time'' program have been summarized by
G\"{u}del et al. (1997), Guinan et al. (2003), Guinan \& Ribas (2004)
and Ribas et al. (2005).

\begin{figure}
\includegraphics[width=4in]{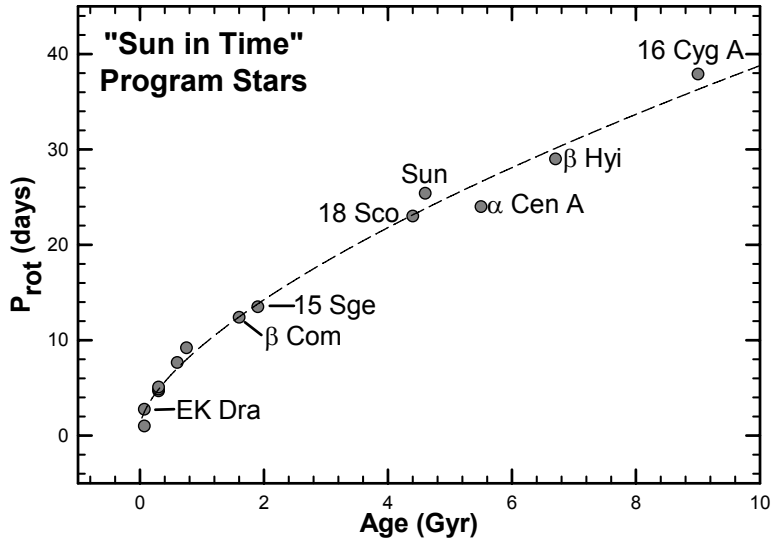}
\vspace{-2mm}
\caption{Plot showing the increase in P$_{\rm rot}$ for dG0--5 stars
with increasing age. This spin-down with age arises from magnetic breaking
from angular momentum loss via magnetized winds.}\label{fig1}
\vspace{-4mm}
\end{figure}

\begin{figure}
\includegraphics[width=4in]{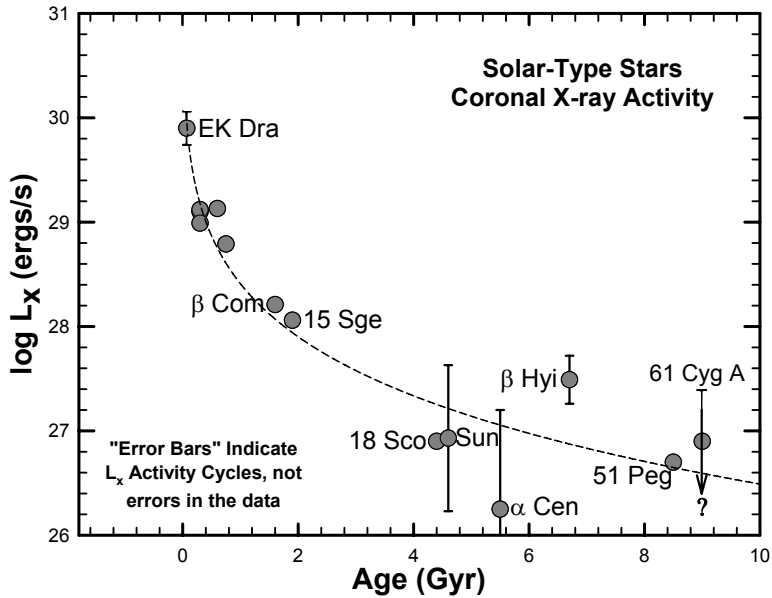}
\vspace{-2mm}
\caption{The variation of X-ray luminosity (L$_{\rm x}$) for solar-type
dG0--5 stars are shown and plotted against age. The ranges in L$_{\rm x}$
for the Sun and other older solar-type stars arise mainly from activity
cycles.}\label{fig2}
\vspace{-8mm}
\end{figure}

The ``Sun in Time'' program also bears on the crucial question of the
influence of the young Sun's strong XUV emissions on the developing
planetary system -- in particular on the photochemical and photoionization
evolution of early planetary atmospheres. The constructed spectral
irradiance tables are of interest to researchers in paleoplanetary
atmospheres and for studies on the atmospheric evolution of the large
number of extrasolar planets found orbiting other G-type stars. The
irradiance data have been published by Ribas et al. (2005) and examples
in the FUV/UV region are given in Fig. 3. As shown, there are dramatic
decreases in the FUV ({\em FUSE}) and FUV/UV ({\em IUE}) emission line
strengths (and irradiances) as the star
loses angular momentum and spins down with age. Fig. 4 shows the averaged XUV
irradiance changes over time for the Sun and other solar-type (dG0--5)
stars from solar
proxies of different ages. The plot illustrates how emissions associated
with hotter plasmas diminish more rapidly as the stars rotate slower with
age. The coronal X-ray/EUV emissions of the young main sequence Sun were
approximately 100--1000x stronger than the present Sun. Similarly,
the TR and chromospheric FUV \& UV emissions of the young
Sun are 10--100x and 5--10x stronger, respectively, than at present.
Also shown in Fig. 4 is the slow increase ($\sim$8\%/Gyr) in the bolometric
luminosity of the Sun over the past 4.6 Gyr. Over this time,
the Sun's luminosity
increased from $\sim$0.7 L$_\odot$ to 1.0 L$_\odot$. This change in solar
luminosity arises from
the acceleration of nuclear fusion in its core that causes its radius and
surface temperature to slowly increase with time. 

\begin{figure}
\includegraphics[width=5in]{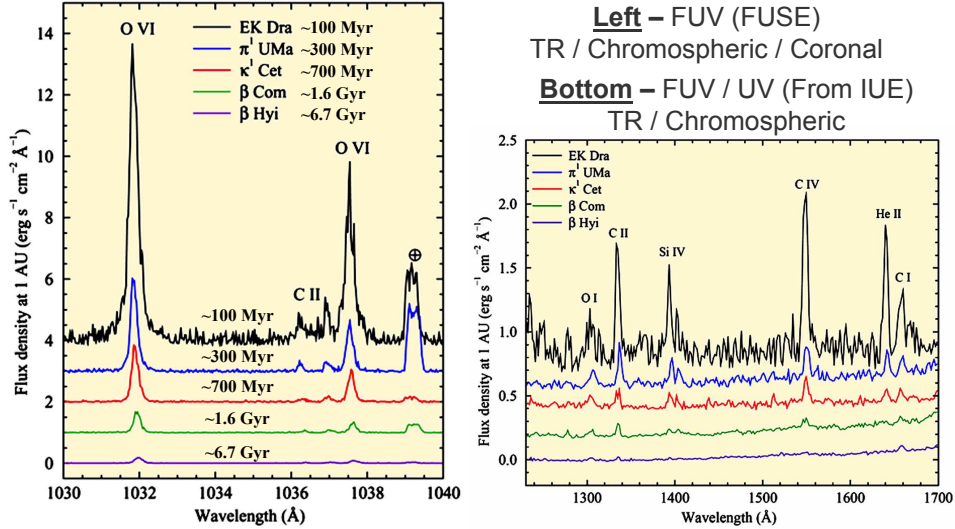}
\vspace{-3mm}
\caption{\small Plots of flux density at 1 AU for the {\em FUSE} O {\sc vi}
(1030--1040\AA) region (left) and the FUV--UV 1200-1700\AA~region from
{\em IUE} low dispersion {\em SWP} spectrophotometry (right). As shown in
the figures
the flux densities decrease rapidly with stellar age. The stars
are plotted from top to bottom in order of increasing age and decreasing
activity.}\label{fig3}
\vspace{-5mm}
\end{figure}

\begin{figure}
\includegraphics[width=4.5in]{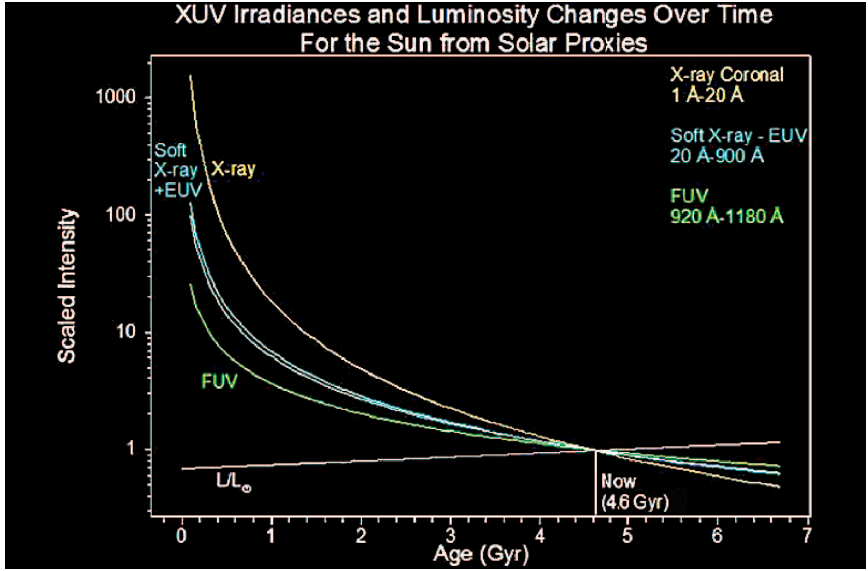}
\vspace{-3mm}
\caption{\small The smoothed XUV spectral irradiances of solar-type stars for
three spectral intervals plotted against age. The flux scales were
normalized to unity at the Sun's age. Also shown is the variation of
bolometric solar luminosity (L/L$_\odot$) over time, where L$_\odot$ = 1 is
the luminosity of the present 4.6 Gyr Sun (L$_\odot$).}\label{fig4}
\vspace{-5mm}
\end{figure}

We have been collaborating with planetary scientists and astrobiology
groups to study the effects of the young Sun's  (and other dG0--5 stars)
strong XUV irradiance on
the loss of water from Mars and its implications for the oxidation of
the Martian soil (Lammer et al. 2003). It has been assumed, from
topographic and geological studies, that Mars was originally warmer and
much wetter than at present, and likely possessed a $\sim$1 bar atmosphere.
Lammer et al. considered ion pick-up sputtering, as
well as dissociative recombination processes. The loss of H$_2$O from
Mars over the
last 3.5 Gyr was estimated to be equivalent to a global martian H$_2$O
ocean with a depth of about $\geq$12 m. If ion momentum transport, a process
to be studied in detail by Mars Express, is significant on Mars, the
water loss may be enhanced by a factor of about 2. For this study it
has been assumed that, for the first billion years, Mars had a hot liquid
iron core and, through rotation, possessed a significant magnetic
field and resulting magnetosphere. This magnetosphere essentially protected
the early Martian environment from the combination of high levels of XUV
radiation and strong winds of the younger Sun that would have otherwise
removed its atmosphere. However, Mars is a smaller, less massive planet
than the Earth, and lost heat at
a faster rate. Thus its iron core solidified $\sim$1 billion
years after the planet's formation. Without the magnetic field, the outer
Martian atmosphere was exposed to the ionizing effects and strong winds
of the early Sun, and thus partially eroded. Photolysis of water ensued,
with a preferential loss of the lighter Hydrogen over the heavier Oxygen.
The loss of water and water vapor from the atmosphere resulted in a greatly
diminished Greenhouse Effect and a rapid cooling of the lower Martian
atmosphere. This rapid cooling (dropping below the freezing point of water)
permitted some water to remain behind, possibly as permafrost trapped below
the Martian soil. This scenario is illustrated in Fig. 5.

A recent study (Kulikov et al. 2006) has been published using
solar-proxy data (from Ribas et al. 2005 and Wood et al. 2002; 2005) to
study the atmosphere and water loss from early Venus. This study shows
that water on early Venus (only 0.71 AU from the Sun) was essentially all
lost during the first $\sim$0.5 Gyr after its formation from the vigorous
action of strong (massive) winds and high XUV fluxes from the young Sun.
In another study, by Grie$\ss$meier et al. (2004),
``Sun in Time'' irradiance date were used to investigate the atmospheric
loss of extrasolar planets resulting from XUV heating, which can lead
to the evaporation of ``Hot Jupiters.''

\begin{figure}
\includegraphics[width=4.5in]{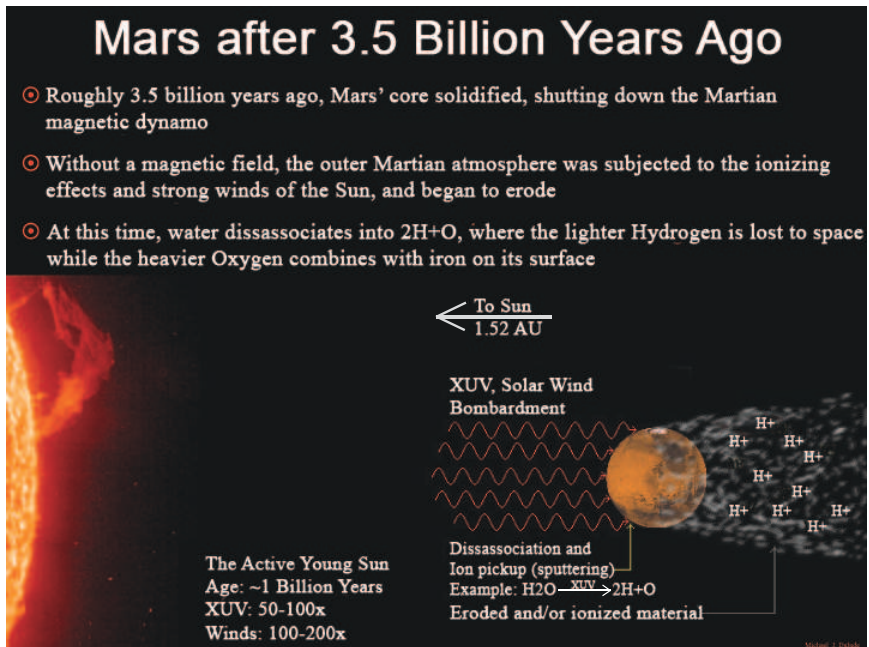}
\vspace{-3mm}
\caption{Cartoon depicting the early evolution and ensuing erosion/loss
of the Martian atmosphere. (Based on Lammer et al. 2003)}\label{fig5}
\vspace{-3mm}
\end{figure}

\vspace{-2mm}
\section{Extension of the ``Sun in Time'' Program to dK--dM Stars}
\vspace{-2mm}

The discovery of extrasolar planets (now 200+ planets) orbiting mostly
main-sequence dG, dK, and dM stars during the last decade has motivated
the expansion of the ``Sun in Time'' program to include cooler, lower
luminosity but very numerous dK and dM stars. Also, with the upcoming
extrasolar planet search and study space missions such as {\em CoRot},
{\em Kepler}, {\em SIM} and, in a decade, {\em Darwin/TPF}, thousands
of additional extrasolar planets are expected to be found. Also, with
{\em CoRot} and {\em Kepler} it will even be possible to discover
Earth-size planets orbiting within the circumstellar liquid water
habitable zones (HZs) of the host stars. 

The goals of this investigation are similar to the original ``Sun
in Time'' program -- {\em 1)} Modeling of dK--dM stars to understand magnetic
activity and dynamo energy generation in low mass stars with very deep
CZs and {\em 2)} Constructing XUV spectral irradiance
tables covering dK--dM stars with a wide range of ages.
The selected program stars ($\sim$15 nearby dM0--5 stars) have well determined
parallaxes, colors,
spectral types and most have observations of age-sensitive measures
such as {\it UVW} space motions, L$_{\rm x}$, Mg~{\sc ii} $h+k$ and 
Ca~{\sc ii} $H+K$ emission fluxes.  

This study, besides improving (and testing) our understanding of
magnetic-related phenomena in cool stars, will help to identify and
characterize dK \& dM stars that have planets suitable for life.
Because of their high space densities, dK stars, and especially dM
stars, will be common targets of extrasolar planet search missions.
Also, their low masses and small radii make them attractive
targets for planet hunting using radial velocity, astrometric, and
transit eclipse methods. dM stars in particular make interesting targets
for exobiology. As shown in Fig. 6, the liquid water circumstellar
HZs for Earth-like planets depend strongly on the
luminosity of the host star (see Kasting \& Catling 2003). The
temperature of the planet depends on the
luminosity of the host star, and the planet's distance (1/r$^2$) from
the star. Also important are the planet's albedo and greenhouse gas heat
trapping contributions. As shown in Fig. 6, because of the low luminosities
of dM stars, their HZs are located quite close to the central star
($\leq$0.4 AU). This makes the hypothetical HZ planet around a dM star more
strongly influenced by stellar flares, winds, and plasma ejection
events (Coronal Mass Ejections = CMEs) that are frequent in young
dM stars (e.g. Kasting et al. 1993; Lammer et al. 2006).
Also, because of the long lifetimes of dM stars ($>$50 Gyr), it might
be possible for life on a planet in the HZ to be much more evolved
(and advanced?) than ourselves. In the following section we focus on
our new program to study the magnetic evolution and XUV fluxes of dM
stars. This program is called ``Living with a Red Dwarf.''








\vspace{-2mm}
\section{``Living with a Red Dwarf'' Program: Magnetic Dynamos, 
FUV \& X-ray Emissions of dM stars}
\vspace{-2mm}

The major aims of the ``Living with a Red Dwarf'' program are to
understand the magnetic 
activity, dynamo structure and plasma physics, as well as determining 
the X-ray/FUV spectral irradiances of dM stars with widely different 
rotations, ages and, therefore, greatly different corresponding levels 
of magnetic activity. The program stars are limited to spectral types
of dM0--5 and cover a wide range of P$_{\rm rot}$ ($<$0.5 -- 200d),
ages ($<$0.1 -- $\sim$13 Gyr), and activity levels (e.g. log L$_{\rm x}$:
$\sim$25.5 -- 29.5 ergs/s). This work complements ongoing related programs
being carried out on main sequence dG--dK stars (e.g. Guinan et
al. 2005; Lakatos et al. 2005). Moreover, this program is a further
important step in
understanding the linkage between magnetic energy generation and
propagation (from stellar dynamo activity) and emission losses in the stellar 
chromosphere, TR, and corona (e.g., Ulmschneider et 
al. 1999; Rammacher \& Cuntz 2003) of stars with very deep convective 
envelopes. As part of this program we are carrying out high precision
photometry to determine rotation periods and starspot coverages. Also, 
we are extensively utilizing NASA/ESA X-ray--UV archival data, and have
several active programs/proposals to obtain further {\em FUSE} and
{\em XMM} observations to fill in the parameter space of
age/rotation/activity for our dM star sample.

For example, the {\em FUSE} FUV spectral region (920--1180\AA) contains
important diagnostic emission lines that can be used to characterize the
energy transfer and atmospheric structure while the ratio of C {\sc iii}
977/1175\AA~fluxes provides valuable information about the TR
electron pressure (see Guinan et al. 2003). These programs have also
become important for 
gauging the FUV emissions of these stars (mostly contributed by the H 
{\sc i} Lyman series, C {\sc iii} 977/1175\AA, and O {\sc vi}
1032/1038\AA~emissions) 
that are critical to assess the photochemical and photoionization 
evolution of planetary atmospheres and ionospheres and the possible 
development of life on extrasolar planets (Kasting 1997; Guinan \& 
Ribas 2002; Kasting \& Catling 2003; Guinan et al. 2003; Ribas et 
al. 2005). 

The characterizations of the XUV irradiances of dM stars 
are important because dM stars comprise $>$70\% of stars in the solar 
neighborhood (and probably in the Galaxy) and have extremely long 
($>$50 Gyr) lifetimes. A fraction of these dM stars may host planets 
within their circumstellar HZs, located nearby 
($\leq$0.4 AU) the host star. On one hand, the frequent flaring of dM stars
(e.g. see G\"{u}del et al. 2003) and possible tidal locking of close-in 
planets (Joshi et al. 1997) could challenge the early development of life 
on a planet located in their shrunken, close-in circumstellar liquid
water HZs. On the other hand, 
were life to form, the longevities of dM stars would be 
favorable to the development of advanced modes of intelligent life.
Although dM stars have not been specifically targeted
in major extrasolar planet searches, so far planets have been
discovered orbiting 3 nearby dM stars (IL Aqr, GJ 581, and GJ 424). In
fact, two of the three planets orbiting IL Aqr are located within the
star's HZ -- one of them is a ``Super-Earth'' (Rivera et al. 2005).
Moreover, theoretical studies by Boss (2006) 
predict the rapid formation of ``Super Earths'' around dM stars indicating
that Earth--like planets hosted by dM stars could be numerous.
More recently, several additional dM stars in the direction of the galactic
bulge have been found, from {\em HST/ACS} photometry, to show transit eclipses
arising from Jupiter-size
planets (Sahu et al. 2006). Interestingly, these extrasolar planets orbit
with ultra-short periods (P$_{\rm orb}$ $<$ 1 day). 
 
Because of their deep outer CZs, dM stars have very
efficient magnetic dynamos and, consequently, f$_{\rm XUV}$/f$_{\rm bol}$ 
values of 10$^2$--10$^4$ higher than solar-type dG stars of comparable 
ages. Thus, possible Earth-like planets orbiting within the shrunken 
HZs near to the dM star host will receive enhanced XUV radiation relative 
to their incident bolometric irradiances. This could have interesting 
and significant photochemical and photoionization effects on the
outer atmospheres of hosted HZ planets. On the other hand, as shown
in Fig. 7, dM stars are cool and have no significant photospheric
UV--NUV fluxes at wavelengths shorter than $\sim$2500\AA. Thus, their NUV
fluxes are far below those of dG and early dK stars. An excellent summary
appraising the habitability of planets hosted by dM stars is given by
Tarter et al. 2006.

\vspace{-2mm}
\section{Initial Results: Nuclear and Magnetic Evolution of dM Stars}
\vspace{-2mm}

Unlike more massive and luminous dG and early dK stars, the nuclear reaction rates in the cores of dM stars
are very slow, resulting in their low luminosities (L $\leq$ 0.06 L$_\odot$)
and extremely long main sequence
lifetimes ($\tau$ $>$ 50 Gyr). As shown in Fig. 8, the luminosities of dM stars are essentially
constant over tens of billions of years after their arrival on the main
sequence. For this reason, their HZs remain fixed over eons of time, ensuring
a stable energy source for a possible hosted planet. By
contrast, stars like the Sun undergo significant changes in luminosity on
timescales of billions of years,
causing their HZs to move slowly outward over time. For example, in $\sim$1
billion
years, our Earth will no longer lie within the then more luminous Sun's HZ,
resulting in the commencement of the runaway Greenhouse Effect on Earth,
in which the oceans begin to evaporate. 

Given their low surface temperatures, dM stars radiate predominantly 
in the near IR, and they have essentially no photospheric continuum flux
$<$ $\sim$2500\AA~(see Fig. 7). However, because of their deep CZs and
resulting efficient magnetic dynamos, dM stars have strong magnetic
dynamo generated coronal/TR/chromospheric FUV/UV emissions with XUV
surface fluxes exceeding those of dG stars with comparable ages.
Similar to the results of the ``Sun in Time'' program, young dM stars
rotate rapidly and have correspondingly strong dynamo-generated XUV
emissions. For example, Fig. 9 shows the decrease in FUV O {\sc vi}
emission strengths between AD Leo (100 Myr) and Proxima Cen (5.8 Gyr).
Young dM stars are also well known for strong and frequent high energy
flares, which emit strong XUV radiation. Moreover,
unlike dG stars, dM stars' magnetic related activity appears to persist
for longer times (a few billion years). Even older dM stars are known
to flare, examples being Proxima Cen ($\sim$5.8 Gyr) and Barnard's Star
($\sim$9 Gyr). A summary of some of the more important characteristics
of dM stars are given below, as well as some implications for planets
orbiting close to them.

\begin{center}
\vspace{2mm}
{\bf Summary of Important dM Star Characteristics and Impacts on Earth-like
planets within their Habitable Zones}
\end{center}

\begin{itemize}
\vspace{-2mm}
\item dM stars have long lifetimes ($>$50 Gyr) and nearly constant
luminosities (see Fig. 8).
\vspace{-2mm}
\item dM stars are ubiquitous, comprising $>$70\% of stars in the
solar neighborhood
and have low masses ($<$0.5 M$_\odot$), temperatures ($<$3900 K) and 
luminosities ($\leq$0.06 L$_\odot$).
\vspace{-2mm}
\item Unlike solar-type stars, dM stars have essentially no photospheric
continua in the UV ($<$2500\AA), because of their low temperatures (see
Fig. 7).
\vspace{-2mm}
\item dM stars have deep outer CZs, very efficient magnetic dynamos
and, consequently, resulting strong coronal X-ray, transition region FUV \&
chromospheric FUV--UV emissions. For example, the ratio of L$_{\rm x}$
to total bolometric luminosity for dM stars is, on average, 10$^3$ times
that of solar-type stars.
(e.g. $[$L$_{\rm x}$/L$_{\rm bol}$$]_{\rm dM}$ $\approx$ 10$^{-3}$ and
$[$L$_{\rm x}$/L$_{\rm bol}$$]_\odot$ $\approx$ 10$^{-6}$)
\vspace{-2mm}
\item The only source of XUV radiation of dM stars is from
dynamo-generated coronal/chromospheric emissions and related flares.
\vspace{-2mm}
\item dM stars have frequent flares even up to relatively old
ages. For example, Proxima Cen ($\tau$ $\approx$ 5.8 Gyr) has about one
large flare per day.
\vspace{-2mm}
\item Theoretical studies by Boss (2006) indicate that ``Super Earths''
can easily form in the proto-planetary disks of dM stars. Planets hosted
by dM stars should be at least as common as those hosted by solar-type
stars. Even without much effort, several dM stars have been found to host
planets (see text).
\vspace{-2mm}
\item dM stars (dM0--5) have HZs located very close to the host star at
$\sim$0.1AU $<$ HZ $\leq$ 0.4AU. (see Fig. 6)
\vspace{-2mm}
\item Although the bolometric irradiance levels within the shrunken HZs
of dM stars would be comparable to those received on Earth from the
present Sun, the dynamo-generated coronal-chromospheric XUV line emissions
at wavelengths
$<$1800\AA~could be 3--10x stronger. This occurs because dM stars have much
higher values of L$_{\rm x}$/L$_{\rm bol}$ and L$_{\rm FUV}$/L$_{\rm bol}$
than the Sun and solar-type stars of 
comparable ages. Thus, because of the proximity of the planet to the dM
star, the X--FUV irradiances are higher.
\vspace{-2mm}
\item dM stars flare frequently and thus emit impulsive XUV 
energies that could be a problem for life on a planet orbiting in the
HZ. However, as shown by Cockell et al. (2000) and
more recently by Cuntz et al. (2006),
even a thin atmosphere, such as that of Mars, does not allow any incoming
FUV/X-ray radiation with wavelengths $<$2000\AA~to reach the surface. On
the other hand, these impulsive bursts of radiation (with $>$2000\AA) could
help, not
hinder the development of life from mutations in the DNA of possible
life forms.
\vspace{-2mm}
\item Although wind and coronal mass ejection events (CME) have yet to be
directly measured for dM stars, scaling from the Sun and inferred wind
properties of younger cool stars from astrosphere Ly-$\alpha$ studies
(see Wood et al. 2002; 2005), it is assumed that young dM stars will have
dense winds enhanced by frequent CME events that could have major effects
on the heating of the exosphere of the planet and on the subsequent possible
erosion and loss of its atmosphere if the planet does not possess a
protective magnetosphere (see Grie$\ss$meier et al. 2004).
\vspace{-2mm}
\item The nearly constant luminosities of dM stars over time scales
of tens of billions of years results in fixed HZs. This provides a
stable environment for life to form and evolve on a possible dM star
HZ planet.  
\vspace{-2mm}
\item In our Galaxy, there are many old dM stars ($\tau$ $>$ 5 Gyr). This could
mean that life (assuming that it formed in the first place ) on a HZ
planet around a dM star could be much more evolved and more advanced
than us at 4.6 Gyr. However, for very old, metal poor, Pop II dM stars
-- like Barnard's
Star, Kapteyn's Star and others, there could be a problem for
terrestrial (rocky) planets to form because of the paucity of metals.
Also, a low metal environment could be problematic for the development
of life.  

\end{itemize}



\begin{figure}
\includegraphics[width=5in]{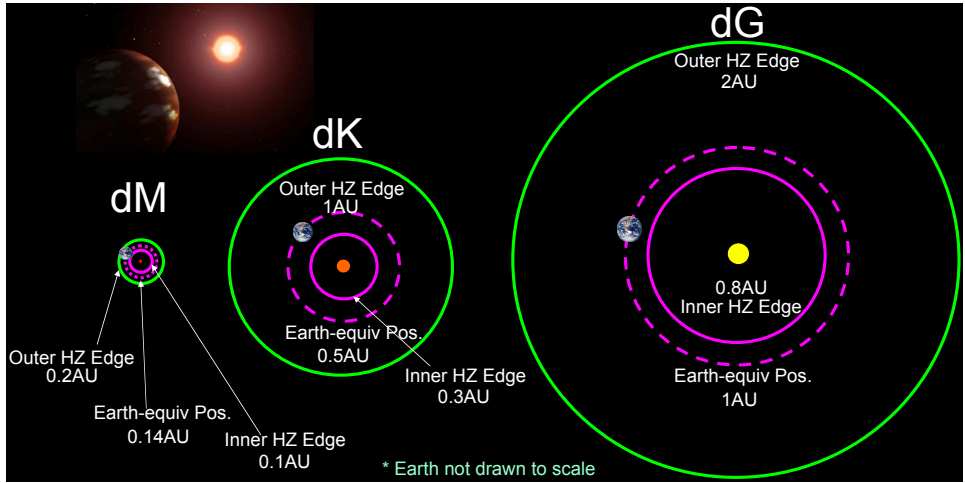}
\vspace{-3mm}
\caption{Approximate circumstellar liquid water
habitable zones for Greenhouse-enhanced Earth-like planets
around dM(dM0--5), dK(dK0--5) \& dG(dG0--5) stars.}\label{fig7}
\vspace{-2mm}
\end{figure}

\begin{figure}
\includegraphics[width=5in]{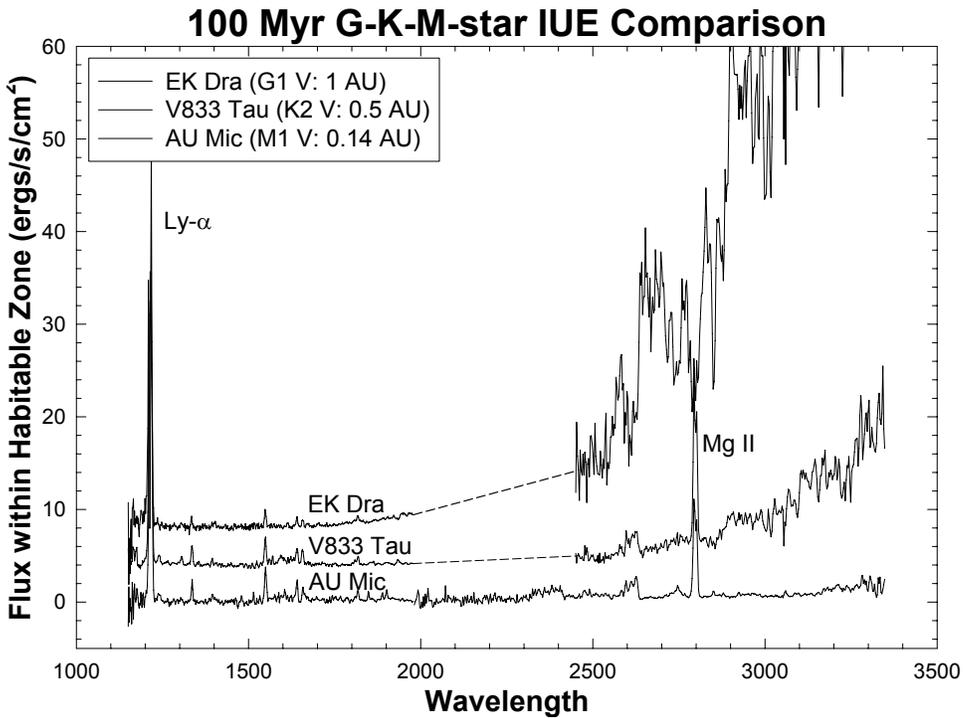}
\caption{FUV--UV fluxes received from young ($\sim$100 Myr) dG, dK
and dM stars at the Earth-equivalent distances within their relative
habitable zones. The data were secured from the {\em IUE} archives.}\label{fig6}
\vspace{-5mm}
\end{figure}

\begin{figure}
\includegraphics[width=5in]{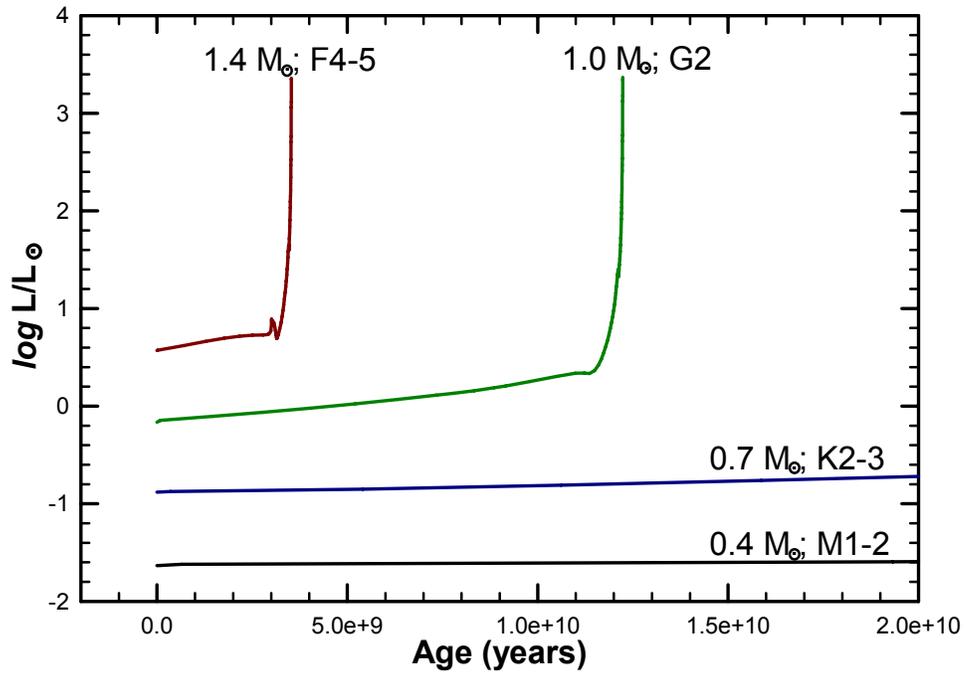}
\vspace{-3mm}
\caption{Stellar evolutionary tracks for 1.4M$_\odot$, 1.0M$_\odot$,
0.7M$_\odot$ and 0.4M$_\odot$ stars. These masses roughly correspond to
dF4--5, dG2, dK2--3 and dM1--2 spectral types, respectively. Data taken
from the Padova Database (http://pleiadi.pd.astro.it/).}\label{fig9}
\vspace{-8mm}
\end{figure}

\begin{figure}
\includegraphics[width=5in]{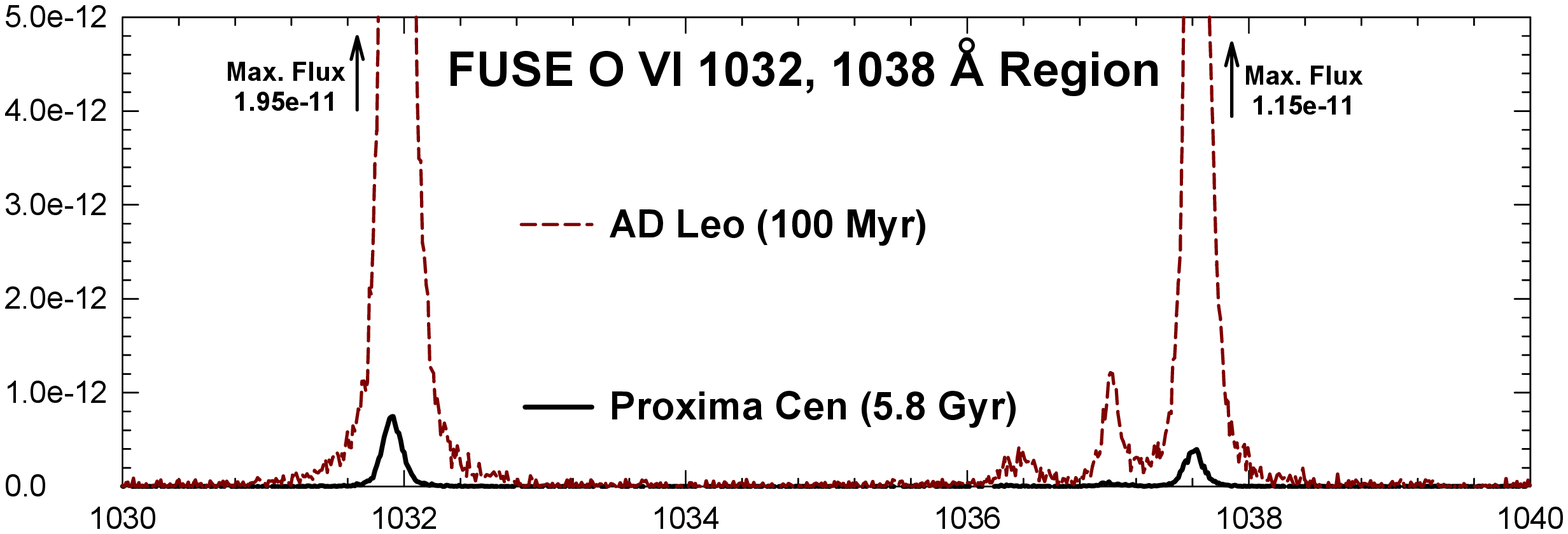}
\vspace{-1mm}
\caption{Plot of the {\em FUSE} O {\sc vi} FUV region for a young (AD Leo;
$\sim$100 Myr) and an old (Proxima Cen; $\sim$5.8 Gyr) dM star,
indicating the large decrease in FUV emissions with age.}\label{fig8}
\vspace{-3mm}
\end{figure}



\noindent
{\it Acknowledgements:}

This research is supported by grants from NASA for data obtained from
{\em FUSE}, {\em HST} and {\em XMM}, as well as an NSF/RUI grant that
supports the photometric observations.


\begin{thebibliography}{}

\bibitem{}
Ayres, T. R.
\newblock{Evolution of the Solar Ionizing Flux},
\newblock{\em Journal of Geophysical Research}, 102, 1641, 1997

\bibitem{}
Boss, A. P.
\newblock{Rapid Formation of Super-Earths around M Dwarf Stars},
\newblock{\em Astrophysical Journal}, 644, 79, 2006

\bibitem{}
Cockell, C. S., Catling, D. C., Davis, W. L., Snook, K., Kepner, R. L.,
Lee, P. and McKay, C. P.
\newblock{The Ultraviolet Environment of Mars: Biological Implications
Past, Present, and Future},
\newblock{\em Icarus}, 146, 343, 2000

\bibitem{}
Cuntz, M., Gurdemir, L., Guinan, E. F. and Kurucz, R. L.
\newblock{Astrobiological Effects of Stellar Radiation in Circumstellar
Environments},
\newblock{\em in press}

\bibitem{}
Grie$\ss$meier, J.-M., Stadelmann, A., Penz, T., and 7 co-authors
\newblock{The Effect of Tidal Locking on the Magnetospheric and
Atmospheric Evolution of ``Hot Jupiters''}
\newblock{\em Astronomy \& Astrophysics}, 425, 753, 2004

\bibitem{}
G\"{u}del, M., Guinan, E. F. and Skinner, S. L.
\newblock{The X-Ray Sun in Time: A Study of the Long-Term Evolution
of Coronae of Solar-Type Stars},
\newblock {\em Astrophysical Journal}, 483, 947, 1997

\bibitem{}
G\"{u}del, M., Audard, M., Kashyap, V. L., Drake, J. J. and Guinan, E. F.
\newblock{Are Coronae of Magnetically Active Stars Heated by Flares? II.
Extreme Ultraviolet and X-ray Flare Statistics and the Differential Emission
Measure Distribution},
\newblock{\em Astrophysical Journal}, 582, 423, 2003

\bibitem{}
Guinan, E. F. and Ribas, I.
\newblock{Our Changing Sun: The Role of Solar Nuclear Evolution
and Magnetic Activity on Earth's Atmosphere and Climate},
\newblock{\em ASPC}, 269, 85, 2002

\bibitem{}
Guinan, E. F., Ribas, I. and Harper, G. M.
\newblock{Far-Ultraviolet Emissions of the Sun in Time: Probing Solar
Magnetic Activity and Effects on Evolution of Paleoplanetary Atmospheres},
\newblock{\em Astrophysical Journal}, 594, 561, 2003

\bibitem{}
Guinan, E. F. and Ribas, I.
\newblock{Evolution of the Solar Magnetic Activity over Time and
Effects on Planetary Atmospheres},
\newblock{\em IAUS}, 219, 423, 2004

\bibitem{}
Guinan, E., Engle, S., Ribas, I. and Harper, G.
\newblock{{\em FUSE} Observations of Young to Old dG, dK \& dM Stars:
Critical Tests of Dynamos, X-FUV irradiances and Impacts on
Planetary Enviroments and the Development of Life},
\newblock{\em BAAS}, 37, 1490, 2005

\bibitem{}
Joshi, M. M., Haberle, R. M. and Reynolds, R. T.
\newblock{Simulations of the Atmospheres of Synchronously Rotating
Terrestrial Planets Orbiting M Dwarfs: Conditions for Atmospheric
Collapse and the Implications for Habitability},
\newblock{\em Icarus}, 129, 450, 1997

\bibitem{}
Kasting, J. F., Whitmore, D. P. and Reynolds, R. T.
\newblock{Habitable Zones Around Main Sequence Stars}
\newblock{\em Icarus}, 101, 108, 1993

\bibitem{}
Kasting, J. F.
\newblock{Habitable Zones around Low Mass Stars and the Search
For Extraterrestrial Life},
\newblock{\em Origins of Life and Evolution of the Biosphere}, 27, 291, 1997

\bibitem{}
Kasting, J. F. and Catling, D.
\newblock{Evolution of a Habitable Planet}
\newblock{\em Annual Review of Astronomy and Astrophysics}, 41, 429, 2003

\bibitem{}
Kulikov, Yu. N., Lammer, H. I. M., Lichtenegger, N. and 8 co-authors
\newblock{Atmospheric and Water Loss from Early Venus},
\newblock{\em Planetary and Space Science}, in press

\bibitem{}
Lakatos, S. L., Voyer, E. N., Guinan, E. F., DeWarf, L. E.,
Ribas, I. and Harper, G. M.
\newblock{Magnetic Activity and High Energy XUV Irradiances of
Dwarf K-Stars - Impacts of XUV Emissions on Hosted Extrasolar Planets},
\newblock{\em BAAS}, 37, 442, 2005

\bibitem{}
Lammer, H., Lichtenegger, H. I. M., Kolb, C., Ribas, I., Guinan, E. F.,
Abart, R. and Bauer, S. J.
\newblock{Loss of Water from Mars: Implications for the Oxidation of the Soil},
\newblock{\em Icarus}, 165, L9, 2003

\bibitem{}
Lammer, H., Lichtenegger, H. I. M., Kulikov, Y. N. and 8 co-authors
\newblock{CME Activity of Low Mass M stars as an Important Factor for
Habitability of Terrestrial Planets. II.},
\newblock{\em Astrobiology}, in press, 2007

\bibitem{}
Rammacher, W. and Cuntz, M.
\newblock{Acoustic Heating Models for the Basal Flux Star tau
Ceti Including Time-dependent Ionization: Results for Ca II
and Mg II Emission},
\newblock{\em Astrophysical Journal}, 594, L51, 2003

\bibitem{}
Ribas, I., Guinan, E. F., G\"{u}del, M. and Audard, M.
\newblock{Evolution of the Solar Activity over Time and Effects on
Planetary Atmospheres. I. High-Energy Irradiances (1-1700\AA)},
\newblock{\em Astrophysical Journal}, 622, 680, 2005

\bibitem{}
Rivera, E. J., Lissauer, J. J., Butler, P. R. and 6 co-authors
\newblock{A $\sim$7.5 M$_\oplus$ Planet Orbiting the Nearby Star, GJ 876},
\newblock{\em Astrophysical Journal}, 634, 625, 2005

\bibitem{}
Sahu, K. C., Casertano, S., Bond, H. E. and 13 co-authors 
\newblock{Transiting Extrasolar Planetary Candidates in the Galactic Bulge},
\newblock{\em Nature}, 443, 534, 2006

\bibitem{}
Tarter, J. C., Backus, P. R., Mancinelli, R. L. and 29 co-authors
\newblock{A Re-appraisal of the Habitability of Planets Around M-Dwarf Stars},
\newblock{\em astro-ph/0609799},2006

\bibitem{}
Ulmschneider, P., Theurer, J., Musielak, Z. E. and Kurucz, R.
\newblock{Acoustic wave energy fluxes for late-type stars. II.
Nonsolar metallicities},
\newblock{\em Astronomy \& Astrophysics}, 347, 243, 1999

\bibitem{}
Wood, B. E., M\"{u}ller, H.-R., Zank, G. P. and Linsky, J. L.
\newblock{Measured Mass-Loss Rates of Solar-like Stars as a Function
of Age and Activity},
\newblock{\em Astrophysical Journal}, 574, 412, 2002

\bibitem{}
Wood, B. E., Redfield, S,. Linsky, J. L., M\"{u}ller,
H.-R. and Zank, G. P.
\newblock{Stellar Ly-$\alpha$ Emission Lines in the {\em Hubble Space
Telescope} Archive: Intrinsic Line Fluxes and Absorption from the
Heliosphere and Astrospheres},
\newblock{\em Astrophysical Journal Supplement Series}, 159, 118, 2005

\end{thebibliography}
\end{document}